%% file: main.tex
\documentclass[acmtog, authorversion, nonacm]{acmart}

\usepackage{booktabs}   
\usepackage{subcaption} 
\usepackage[utf8]{inputenc} 
\usepackage[T1]{fontenc}    
\usepackage{url}            
\usepackage{amsfonts}       
\usepackage{nicefrac}       
\usepackage{microtype}      
\usepackage{graphicx}
\usepackage{wrapfig}
\usepackage{multirow}
\usepackage{xcolor}
\usepackage{algorithmic}
\usepackage{paralist}
\usepackage{comment}
\usepackage{enumitem}
\usepackage[english]{babel}
\usepackage{epstopdf}
\usepackage{lipsum}
\usepackage{ mathrsfs }
\usepackage{cancel}

\newcommand{\scheme}{\textsc{SpecSyn}~}

\begin{document}



\title{Large Language Models Based Automatic Synthesis of Software Specifications} 


\author{Shantanu Mandal}
\affiliation{%
  \institution{Texas A\&M University}
  \country{USA}
  }
\email{shanto@tamu.edu}

\author{Adhrik Chethan}
\affiliation{%
  \institution{Texas A\&M University}
  \country{USA}
}

\author{Vahid Janfaza}
\affiliation{%
  \institution{Texas A\&M University}
  \country{USA}
}

\author{S M Farabi Mahmud}
\affiliation{%
  \institution{Texas A\&M University}
  \country{USA}
}

\author{Todd A Anderson}
\affiliation{%
  \institution{Intel Labs}
  \country{USA}
}

\author{Javier Turek}
\affiliation{%
  \institution{Intel Labs}
  \country{USA}
}

\author{Jesmin Jahan Tithi}
\affiliation{%
  \institution{Intel Labs}
  \country{USA}
}

\author{Abdullah Muzahid}
\affiliation{%
  \institution{Texas A\&M University}
  \country{USA}
  }

\renewcommand{\shortauthors}{Mandal, et al.}

\input{sub_abs}






\maketitle


\section{Introduction}
\label{sec-intro}
\input{sub_intro}

\section{\scheme: Specification Synthesis Framework}
\label{sec-main}
\input{sub_desc.tex}

\section{Results}
\label{sec-results}
\input{sub_results.tex}

\section{Related Work}
\label{sec-rel}
\input{sub_rel.tex}

\section{Discussion and Future Work}
\label{sec-future}
\input{sub_future.tex}

\section{Conclusion}
\label{sec-conc}
\input{sub_conc}

\bibliography{main.bib}
\bibliographystyle{ACM-Reference-Format}


\end{document}

%% file: sub_abs.tex
\begin{abstract}
    
Software configurations play a crucial role in determining the behavior of software systems. In order to ensure safe and error-free operation, it is necessary to identify the correct configuration, along with their valid bounds and rules, which are commonly referred to as software specifications. As software systems grow in complexity and scale, the number of configurations and associated specifications required to ensure the correct operation can become large and prohibitively difficult to manipulate manually. Due to the fast pace of software development, it is often the case that correct software specifications are not thoroughly checked or validated within the software itself. Rather, they are frequently discussed and documented in a variety of external sources, including software manuals, code comments, and online discussion forums. Therefore, it is hard for the system administrator
to know the correct specifications of configurations due to the lack of clarity, organization, and a centralized unified source to look at.
To address this challenge, we propose \scheme\, a framework that leverages a state-of-the-art large language model 
to automatically synthesize software specifications from natural language sources. Our approach formulates software specification synthesis as a sequence-to-sequence learning problem and investigates the extraction of specifications from large contextual texts. This is the {\bf first} work that uses a large language model for end-to-end specification synthesis from natural language texts. Empirical results demonstrate that our system outperforms prior state-of-the-art specification synthesis tool
by 21\% in terms of F1 score and can find specifications from single as well as multiple sentences. 

\end{abstract}

%% file: sub_intro.tex
Software configurations represent an essential component of software systems. System failures induced by software misconfigurations (i.e., not setting various configuration parameters according certain specifications) have become increasingly common 
~\cite{how_hadoop_cluster_break, yin.osp.2011, gunawi.cc.2016}. Such misconfigurations give rise to a range of issues, including application outages, security vulnerabilities, and inaccuracies in program execution~\cite{xu.ecse.2015,eshete2011early,DBLP:journals/csur/XuZ15}. 
The adverse impact of software misconfigurations is demonstrated in several high-profile cases~\cite{269497,cnn.2017,tc.2019,fw.2017}. 
For instance, a configuration error caused by an internet backbone company, resulted in a nationwide network outage in 2017~\cite{cnn.2017}. Similarly, in 2019, millions of Facebook users were affected by a server misconfiguration outage~\cite{tc.2019}. Furthermore, a system configuration change led to a five-hour outage of AT\&T's 911 service, preventing numerous callers from accessing the emergency line~\cite{fw.2017}.
Therefore, the development of effective tools to help prevent
software misconfigurations is of utmost importance.

The research community has recognized the criticality of software misconfiguration and proposed numerous efforts to address it by developing techniques to check, troubleshoot and diagnose configuration errors~\cite{267138, 10.1145/1294261, DBLP:journals/csur/XuZ15}. However, the  majority of the approaches can generally only be applied after an error has occurred. Primarily, the root cause of software misconfiguration is attributed to human errors. Thus, a more effective strategy for avoiding such misconfigurations would be to guide or enforce the correct usage of configuration practices, rather than identifying and correcting them after a failure has already occurred.
Typically, software configurations are set by software administrators. To aid these administrators, software vendors often release user manuals that describe different configuration specifications. These manuals, typically available in PDF or HTML format, provide guidance on the correct and recommended setup of configurations to system administrators, containing detailed textual descriptions of configuration parameters, their descriptions, usage, and constraints. Nevertheless, as these manuals are exceedingly voluminous, many administrators tend not to read them in detail and instead rely on intuition to configure software~\cite{xu.sosp.2013, novick.sigdoc.2006},
frequently leading to misconfiguration and subsequent software failures. Hence, it is crucial to develop an automatic tool that extracts configuration specifications from these sources, to provide guidance to administrators, or integrate automated tools that suggest best practices. Thus, this paper investigates the feasibility of building an automatic tool capable of extracting specifications from unstructured sources of configuration descriptions, which are predominantly written in a natural language such as English.

Specifications refer to a set of legitimate rules or guidelines that dictate the configuration of software. Failure to adhere to these specifications by configuring the software with invalid parameters may result in software malfunction. Extensive research has been conducted to extract software specifications from unstructured specification sources~\cite{pracExtractor,configV,nadi.icse.2014}. For instance, in PracExtractor~\cite{pracExtractor}, Xiang et al. utilize the Universal Dependency algorithm~\cite{de-marneffe-etal-2014-universal} to synthesize specifications from software manuals. This 
technique establishes a syntactic mapping between various parts of speech within a sentence. To construct the set of syntactic relationship trees, they initially collected samples from software manuals that contain valid specifications. They then attempted to match other sentences' syntactic relation with the collected syntactic relation tree. If a match was found, it implied that the samples also contained specifications and it was extracted based on the relation. The authors of ConfigV~\cite{configV} have employed a rule-based approach to synthesize specifications from configuration files. This approach involves initially parsing a training set of configuration files, which may be partially correct, to create a well-structured and probabilistically-typed intermediate representation. A learner that utilizes an association rule algorithm is then employed to translate this intermediate representation into a set of rules. These rules are subsequently refined and ranked through rule graph analysis to synthesize specifications. Researchers have also tried to synthesize specifications from programming source code by employing static analysis~\cite{nadi.icse.2014}. However, specifications collected through this approach need more analysis and expert knowledge for refinement by humans. 

The majority of prior methods for synthesizing specifications from unstructured sources have relied on rule-based approaches. However, such approaches are known to have limited generalizability and may require human intervention during certain steps of the synthesis process. Alternatively, learning-based approaches are better suited for discovering relationships in unstructured data, particularly those utilizing deep learning techniques that have shown significant success in various unstructured data domains, including natural languages, images, and videos. As the main objective of this paper is to synthesize specifications from a natural language source, we explore the potential of deep learning-based approaches to address this problem.

To this end, we have framed the specification synthesis problem as an {\bf end-to-end sequence-to-sequence} learning problem. The resulting system is referred to as {\bf \scheme},
which takes texts as an input and produces specifications as an output. The synthesis process involves two steps of prediction. First, it checks whether the input texts contain any specification. If they do, secondly, \scheme\ performs end-to-end synthesis of the specifications. Given that the input for specification synthesis is in the form of a natural language text, we have incorporated a large pre-trained language model to enhance the model's understanding of the context. One such widely used model is BERT~\cite{bert}, a deep learning-based model that pre-trains on a large corpus of English texts to learn the latent representations (i.e., a vector) of words and sentences within their respective contexts. We have fine-tuned~\cite{fine_tune} the BERT model for our specification synthesis task using a custom decoder (a decoder is a part of the language model that produces outputs based on the latent representations). 
The incorporation of this large language model has enabled us to synthesize not only simple single-sentence specifications but also complex sets of specifications from texts consisting of multiple sentences and parameters. Figure~\ref{fig:overview} shows the high level workflow of \scheme.

\subsection{Contributions}
We make the following technical contributions in this paper.

{\bf Problem Formulation:} We formulate the task of software specification extraction from natural language texts as an end-to-end sequence-to-sequence learning problem. This approach involves mapping an input sequence, consisting of natural language texts, to an output sequence that comprises of single or a set of relevant software specifications. 
    
{\bf Contextual Model Integration:} In order to achieve accurate and effective extraction of software specifications from natural language texts, we propose to use the state-of-the-art BERT~\cite{bert} model. By leveraging the acquired knowledge and advanced language processing capabilities of a pre-trained BERT model, we aim to extract relevant specifications from the text in a manner that is both accurate and efficient. Through empirical analysis, we demonstrate the effectiveness of \scheme\ in the context of software specification extraction from natural language texts. To the best of our knowledge, this is the {\bf first} paper that uses a large language model for end-to-end synthesis of configurations specifications from natural language texts.

{\bf Complex Dependency Modeling:} The current investigation presents a model that is able to process
text
consisting of multiple sentences. Specifically, our proposed model is designed to effectively capture complex specification relations within 
longer text.
Notably, the model is capable of discerning the relationships between multiple specifications contained within a single sentence, as well as extracting individual specifications that are connected in a meaningful manner
within a text.
The efficacy of our model is demonstrated through empirical analyses, which provide evidence of its ability to accurately identify and extract relevant information from 
longer text data.

{\bf Generality:} The framework, \scheme, is capable of processing any natural language text, thereby rendering it independent of the source of textual data. Consequently, it does not solely rely on software manuals for the extraction of software specifications. Rather, it can be utilized to extract relevant specifications from a wide range of sources including software codebase comments, as well as online resources such as StackOverflow and discussion forums. This flexibility allows for the construction of a more comprehensive and robust set of specifications, thereby enhancing the overall effectiveness of the framework.


\subsection{Outline}
The remainder of this paper is organized as follows. Section~\ref{sec-main} describes the details of \scheme\ framework. First, the section lays out the specification definition, followed by discussion on specification extraction types and specification categories. Then, different specification sources and dataset construction approaches are described. After that, we describe the model development and integration of a large language model with our framework. Section~\ref{sec-results}
describes the experimental setup and experimental results. Section~\ref{sec-rel}
introduces the background and related work. Section~\ref{sec-future}
discusses potential future work. Finally, Section~\ref{sec-conc} concludes
this paper.

%% file: sub_desc.tex
\begin{figure*}[htbp]
    \begin{center}
        \includegraphics[width=0.9\textwidth]{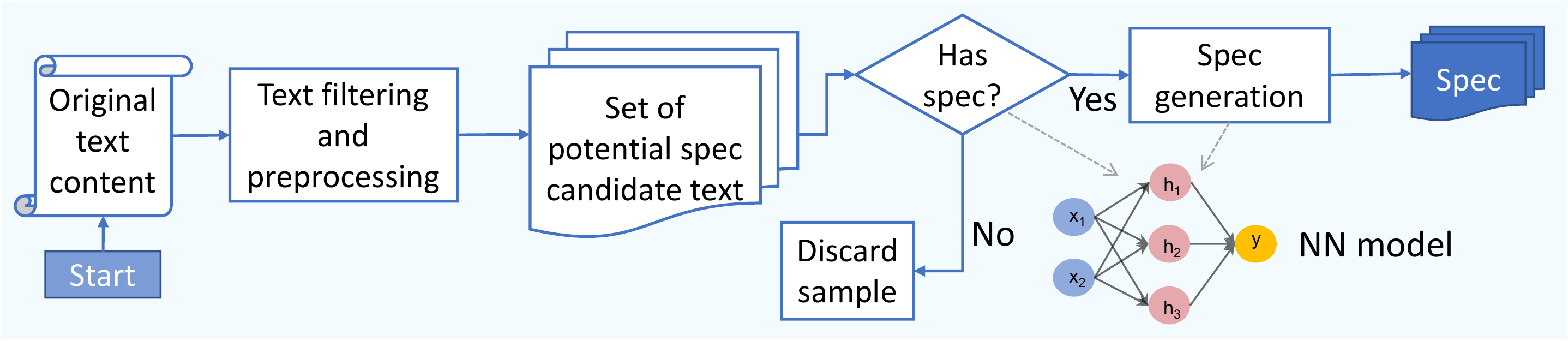}
        \caption{Overview of \scheme}
        \label{fig:overview}
    \end{center}
\end{figure*}

\subsection{Software Specification}

Software specification is a set of rules that consists of relation between various keywords, numbers or pre-formatted string. Here, keywords represent different configuration parameters.
Specification defines how a configuration should be presented by outlining the specific rules and requirements for configuration format and structure. 
Formally, a specification can be defined as Definition 1.

{\bf Definition 1:}    
A specification $S$ is defined as $S = \{R_i\}$, where 
$R_i$ is a rule represented by a tuple, $R_i=\langle K_i, V_i, L_i \rangle$.
Here, \\ 
$K_i \in  \textit{Keyword} $ \\
$V_i \in \{\mathbb{R}, Keyword, S\}~where~\mathbb{R} = Set~of~Real~Numbers, S=String$ \\
$L_i \in \{\emptyset, =, \neq, >, <, \textit{AND}, \textit{OR}, \textit{Interval}, \textit{Set}, \textit{Use}, \textit{With}, \textit{String Format} \} $\\


The goal of \scheme\ is to analyze the natural language texts from various sources and produce a specification, if there is any in the texts. Next we are going to classify types of specification extraction as well as various categories of specifications.


\subsubsection{Specification Extraction Type:} 

Based on the presence of specification in the text, we categorize specification extractions 
into two major types: {\it Simple} and {\it Complex}.
By distinguishing between these two types, we are able to better evaluate the performance of our extraction process. Previous research has focused exclusively on the {\it Simple} extraction category, while the {\it Complex} category demands more sophisticated extraction process and modeling techniques. Table~\ref{tab:extraction_type} shows examples of text for different specification extraction types.


\begin{table}[htbp]
\scalebox{.95}{
\begin{tabular}{p{0.18\columnwidth}p{0.8\columnwidth}}
\toprule
\textbf{Type} & \textbf{Example} \\
\midrule
 Simple & It is necessary to use a number greater than 1500 for $user\_port$\\ \hline
 Complex\textsubscript{Single} & The default pointer size in bytes is used when $max\_rows$ option is specified. This variable should be between 2 and 7.\\ \hline
 Complex\textsubscript{Multi}  & $have\_ssl$ and $have\_open\_ssl$ need to be set True to enable secured connection.\\
 \bottomrule
\end{tabular}
}
\vspace{0.5cm}
\caption{Examples of specification extraction types}
\label{tab:extraction_type}
\end{table}

{\bf Simple Extraction:} 
{\it Simple} extraction refers to specifications extraction process where the specification is located within a single sentence containing only one specification. The majority of specification extractions fall within this category.
The task of modeling {\it Simple} extractions is comparatively less intricate due to the concise nature of general sentences. The specifications are also defined in a more straightforward manner. We observed that fewer training examples are required to train a model for extracting {\it Simple} categories as opposed to {\it Complex} categories.

{\bf Complex Extraction:} 
The second type, known as {\it Complex} Extraction, consists of cases where multiple sentences are required to extract specifications from the text or when there are multiple specifications intertwined in a text that need to be extracted. 
{\it Complex} extraction has been further subdivided into two distinct subcategories, namely {\it Complex\textsubscript{Single}} and {\it Complex\textsubscript{Multi}}. 
The {\it Complex\textsubscript{Single}} category is applicable when only one specification is present in the text, but the extraction process requires examining multiple sentences. 
Typically, in such cases, the first sentence identifies the parameter keyword, while the subsequent sentences describe the specification. The sentences usually connected by some pronoun. 
In order to extract the specification and identify the keyword to which it refers, all sentences are necessary to look at. 
On the other hand, {\it Complex\textsubscript{Multi}} refers to situations where multiple specifications are defined within a text, and multiple parameter keywords are used to refer to a single specification. Basically in this category multiple specifications of multiple keywords are described in a compact intertwined fashion in a natural language.

\begin{table}[htbp]
\begin{tabular}{p{0.18\columnwidth}p{0.8\columnwidth}}
\toprule

\textbf{Category} & \textbf{Example} \\ 
\midrule
 Quantitative & It is recommended to raise the \textit{ulimit} to 10,000, but more likely 10,240
                 because the value is usually expressed in multiples of 1024. \\ \hline 
 Utilization & Mount option \textit{sync} is strongly recommended since it can minimize or avoid reordered writes,
 which results in more predictable throughput.\\ \hline 
 Interrelation & If you are having problems with the service, it is suggested you follow the instructions below  to try starting \textit{httpd.exe} from a console window, and work out the errors before struggling to start it as a service again. \\ \hline 
 Attribute & To avoid the ambiguity, users can specify the plugin option as \textit{--pluginsql-mode}. Use of the \textit{--plugin} prefix for plugin options is recommended to avoid any question of ambiguity.  \\ \hline 
 Generic & It is recommended but not required that \textit{--ssl-ca} also be specified so that the public  certificate provided by the server can be verified.  \\  
 \bottomrule
\end{tabular}
\vspace{0.5cm}
\caption{Different categories of specifications with examples}
\label{tab:spec_cat}
\end{table}

\subsubsection{Specification Categories:}  

We categorize the specification into five major categories based on the underlying specification definition and property.
Among them, {\it Quantitative} represents specifications that are quantifiable, such as numerical or boolean comparisons. {\it Utilization} categorizes any usage of keyword for any particular task, {\it Interrelation} categorizes correlation between keywords, {\it Attribute} categorizes different attribute such as path, domain etc., and {\it Generic} categorizes general suggestions without any specific criteria.
The categories and their examples are described in table~\ref{tab:spec_cat}.

The majority of the specifications identified in our study can be categorized as {\it Quantitative}.
{\it Quantitative} specification can have multiple definitions. The {\it Generic} category lacks a specific definition and is primarily characterized by suggestions or recommendations. The remaining three categories are infrequent and each have only one specification definition. Table~\ref{tab:spec_pat} describes the definition and pattern of each categories.

The significance of specification categorization lies in its potential to facilitate more effective utilization of detected specifications during later stages.
Even though in our work the initial detection of specifications is a binary prediction process independent of their categories, the ability to categorize specifications can provide valuable insights for optimizing their deployment in later stages such as suggesting them to system administrators or incorporating them to other tools.
Therefore, in our work, we also demonstrate the capability of our framework to accurately predict specification categories with different decoders.
Although these categories are intuitive and general to notice, we are inspired about them from prior work~\cite{pracExtractor}.
However, previous research has not explored the detection capabilities of various categories of specification as we have done in our framework.

\begin{table}[hbtp]
\begin{tabular}{p{0.2\columnwidth}p{0.4\columnwidth}p{0.3\columnwidth}}
\toprule
\textbf{Category} &  \textbf{Definition} & \textbf{Pattern} \\
\midrule
 Quantitative & p == v, 
                p < v | p > v,  
                p $\in$ [v, v'], 
                p $\in$ \{v, v'\}, 
                & v\textsubscript{<value>},  
 less\textsubscript{syn} | more\textsubscript{syn} than v\textsubscript{value},  
 between\textsubscript{syn} v'\textsubscript{<value>} to v2\textsubscript{<value>},  v\textsubscript{<value>} or v'\textsubscript{<value>}\\ \hline
 Utilization & use(p) & used\textsubscript{syn} | useful\textsubscript{syn} \\ \hline
 Interrelation & with(p, p'),  
                 prefer(p, p') & along\textsubscript{syn} with p'\textsubscript{para}
                                            prefer\textsubscript{syn}p'\textsubscript{para}\\ \hline
 Attribute & format(p,f) & f\textsubscript{<format>} \\  \hline
 Generic & \{recommended, prioritize, ...\}  \\
                               
 \bottomrule
\end{tabular}
\vspace{0.5cm}
\caption{Specification definition and patterns}
\label{tab:spec_pat}
\end{table}

\subsection{Data Collection}

Data is the lifeline of any deep learning-based solutions, and collecting data to train and test any deep learning-based system is one of the most significant and challenging tasks. For specification synthesis, it becomes even more difficult since it requires highly specific and domain-dependent data. Since there is no standard dataset available for specification synthesis, we have to collect and create our own datasets, which is a time-consuming and resource-intensive task. However, despite the challenges, there are several sources where software specifications can be found. 
The main source of specifications is the software manual. However, it can also be derived from comments embedded in the software code base, particularly when the source code is publicly available. It may be obtained from various other sources also such as Stack Overflow, online discussion forums, and other community-driven platforms. In the following section, we will describe different sources of software specifications and the specification collection process.

\begin{table}[hbtp]
\scalebox{.90}{
\begin{tabular}{lllcc}
\toprule
\textbf{Name} & \textbf{Software type} & \textbf{Format} & \textbf{Pages} & \textbf{Keyword} \\
\midrule
MySQL & Database & PDF & 6644 & Yes\\ \hline
PostgreSQL & Database & PDF & 3055 & Yes\\ \hline
 
HDFS & Distributed Storage & HTML & 2331 & Yes\\ \hline
HBase & Distributed Storage & HTML & 787 & Yes\\ \hline
Cassandra & Distributed Storage & HTML & 50 & No\\ \hline
Spark & Distributed Computing & PDF & 66 & Yes\\ \hline

HTTPD & Web Server & HTML & 1009 & Yes\\ \hline
NGINX & Proxy & HTML & 900 & No\\ \hline
Squid & Proxy & HTML & 330 & Yes\\ \hline
Flink & Stream Processing & HTML & 1434 & Yes\\ \hline

\end{tabular}
}
\vspace{0.5cm}
\caption{Description of software manuals}
\label{tab_data_man_desc}
\end{table}

\subsubsection{Data Source:}
We mainly collect specifications from three major sources. The most important source of the specification is software manuals. The major portion of specifications is collected from manuals. We also collected specifications from source code comments and other online sources.  

{\bf Software Manuals}
The main source of software specification is the software manual. These manuals are typically provided by the corresponding software vendor and contain detailed information about how to use the software, including its features, functions, and limitations. Software manuals are often available in different formats such as PDF, HTML, or in online. Table~\ref{tab_data_man_desc} details the summary of software manuals that we used to collect specifications.
In order to construct our dataset, we gathered specifications from 10 diverse software manuals, spanning a range of software domains.
Typically, software manuals are extensive in nature, containing a considerable amount of information. For instance, MySQL's manuals consist of a total of around 6500 pages. Due to the sheer volume of information contained in these manuals, it is impractical to read them line by line. Therefore, a keyword-based filtering method is employed to extract only the relevant sentences for closer examination. These keywords typically correspond to configuration parameters, which can be located within the manuals. 
The majority of software packages typically have their keywords listed either in the manuals or on their configuration page. However, in our study, we were unable to find such a listing for NGINX and Cassandra.
In the case of these two software packages, we solicited human assistance to aid us in extracting the specifications.
While simple specifications can often be derived from individual sentences, more complex specifications may require analysis of neighboring sentences. As complex specifications may be embedded within the text, the focus is on the section of the text where the keyword is present.

{\bf Other Sources} 
In addition to the manual-based specification,
we also collected specifications from two other additional sources: software code base comments and online discussion forums. As our method operates with natural language, it is independent to the source of the data, provided that it is composed in a natural language. The inclusion of these alternative sources is primarily intended to demonstrate the generalizability of our framework across a broad range of text-based specification descriptions.

To demonstrate the source code base specification generalizability, we develop and integrate a parser into our framework and apply the parsing to MySQL source code only. However, the same methodology can be applied to parsing other software as well. For the purpose of parsing comments from MySQL source code, we use a Python-based parser. It is essential to be cautious when parsing the source code in this manner, as the codebase also contains commented-out codes. But we are only interested in text-based comments. Therefore, commented out code needs to be discarded for a better-refined dataset. In addition, a crawler is also developed to extract software keyword-specific posts from StackOverflow to augment the specification-related dataset. 

\begin{table}[hbtp]
\scalebox{.93}{
\begin{tabular}{ll}
\toprule
\textbf{Tag} & \textbf{Pattern} \\
\midrule
 <bool> & "enable" | "on" | "true" | "disable" | "false" | "off" | ...\\ \hline
 <num> & $\forall$ w $\in$ $\mathbb{R}$  \\ \hline 
 <unit> & “byte” | “MB” | “ms” | “\%” | ...  \\ \hline  
 <keyword> & $\forall$ w $\in$ Configuration Parameter Name  \\ \hline
 <format> & “email address” | “absolute path” | “domain name” | ...\\ \hline
\end{tabular}
}
\vspace{0.5cm}
\caption{Description of pattern for data composition}
\label{tab:spec_pattern}
\end{table}

\subsubsection{Data Composition:}

Prior to generating the candidate specification texts from the original contents, our system refines the dataset through a series of preprocessing steps. First, the texts are divided into smaller candidate texts based on the extraction type. Then it verifies the presence of the keyword in the candidate texts. If the keyword is absent, the candidate texts are discarded and not considered as a potential sample for further processing. 
Both the extraction types (i.e., {\it Simple} and {\it Complex}) require the presence of relevant keywords in the candidate texts. The keywords can easily be found for each of the software in different places (e.g., manual, web).
Upon identifying potential candidates, the system proceeds to search for predefined patterns within the candidate text as specified in Table~\ref{tab:spec_pattern}. These patterns are then replaced with corresponding tags. In instances where identical patterns are present multiple times within the text, tags are differentiated through the use of different identifier. 
This process enhances the generality of the potential candidates, thereby enabling easier detection and synthesis of specifications by the model.
Thus, each candidate comprises of tuple $\langle C, T \rangle$, where $C$ represents the candidate text and $T$ represents the associated pattern tag set. After detecting and synthesizing specification based on $C$, the system can reconstruct the original representation of the specification by replacing the tag in synthesized specification of $C$ with the corresponding pattern in $T$.

One may argue that 
passing the keyword set for finding potential specification candidate text may introduce more redundancy. Rather, the system should identify the keyword itself. However, knowing the keyword set for a particular software is necessary. Multiple software can have same keyword with different specification rule. Therefore, if it is not known for which software the specifications are being synthesized, then the system can synthesize a syntactically correct but potentially wrong specification for a software. This is especially true 
if \scheme\ synthesizes specification from other sources.
Therefore, knowing for which software the specifications are being synthesized and corresponding keyword set is quite important. 
Also, Keyword-based filtering enables us to accomplish two goals. First, it eliminates a significant portion of the samples that are irrelevant in nature. 
Since a model can easily recognize these and accurately classify them as non-specification, the model's performance would be heavily skewed towards making accurate predictions of true negatives. Therefore, considering sentences that has keywords or solely considering neighboring sentences with keywords will allow for a fairer performance comparison.
Second, keyword based filtering also discards a major portion of false positives.
For example text like “See page 157 for details of MySQL 11.7.8” does not hold any specification, but this can be detected as specification if the model is not trained with larger number of samples.
In our study, 
we discovered that a model can also be trained for all cases even without implementing keyword-based filtering. However, such approach requires a larger number of samples to be used for model training. 
Also, identifying syntactically valid but semantically incorrect specifications requires additional failure checks through software execution.
Therefore, due to resource and time constraint, in this project, we pursue a keyword-based filtering process and keep the other ideas as potential future work.

\subsection{Model Development}

\subsubsection{Contextual Model Integration: BERT}
Contextual model integration refers to the process of combining language models to improve the performance of natural language processing tasks. The idea is to leverage the strengths of different models and combine them in a way that captures the complex relationships between words and their contexts in a given text.
One of the approaches of contextual model integration is to use a hierarchical model, where one model is used to capture the overall context of the input sequence and another model is used to make more specific predictions based on the context.
In our case, we use BERT (Bidirectional Encoder Representations from Transformers), a pre-trained large language model~\cite{bert}, that is trained on a large dataset of natural language texts and we fine-tune it with our task-specific custom decoder.

BERT~\cite{bert} is a powerful neural network model that has revolutionized the field of natural language processing (NLP) in recent years. 
It was first introduced by Google in 2018 and has since become one of the most widely used and effective language models in NLP. The core idea behind BERT is to leverage the power of Transformer-based architectures~\cite{transformer} to create a deep bidirectional language model that can capture contextual information from both directions of the input sequence. Unlike traditional language models, which are trained in a left-to-right or right-to-left fashion, BERT is trained using a masked language model (MLM) objective that randomly masks certain tokens in the input sequence and requires the model to predict the missing word based on the surrounding context. 
One of the key innovations of BERT is the use of multiple layers of self-attention to capture complex relationships between words in the input sequence. Each layer of the model contains a self-attention~\cite{transformer} mechanism that allows the model to attend to different parts of the input sequence and capture dependencies between words that are far apart in the input. 
Additionally, BERT uses a combination of word embeddings, positional embeddings, and segment embeddings to capture both the meaning and position of words in the input sequence. All of these embeddings are self-learned through back-propagation while training on a larger dataset. BERT has proven to be highly effective for a wide range of NLP tasks, including text classification, question-answering, language translation, etc .~\cite {Qiu2020}. In order to adapt BERT to specific tasks, researchers typically fine-tune~\cite{fine_tune} the model by adding a task-specific output layer and training the model on a task-specific dataset. The fine-tuning process allows the model to learn task-specific features and improve its performance on the target task.

\begin{figure}[htpb]
    \centering
    
    \begin{subfigure}{.9\columnwidth}
        \includegraphics[width=\textwidth]{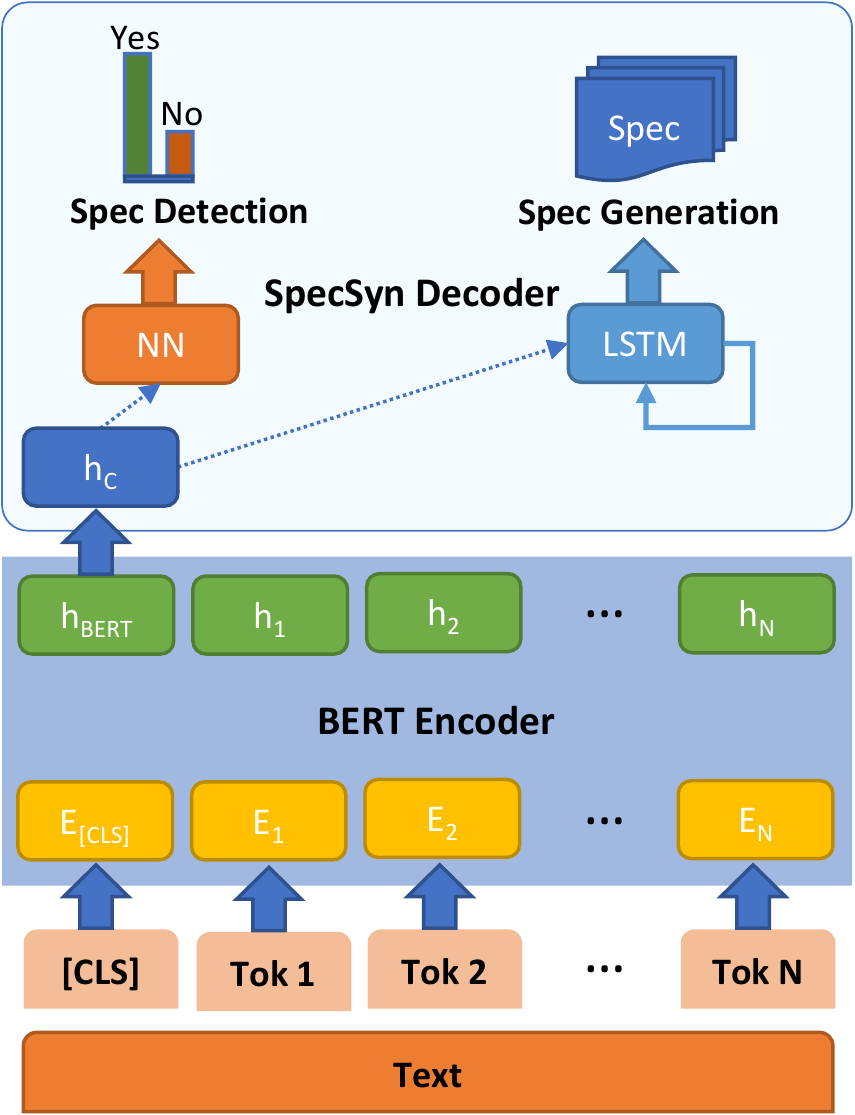}
    \end{subfigure}

    \caption{\scheme\ model architecture}
    \label{fig:model}
\end{figure}

\subsubsection{Specification Detection and Generation}

The specification synthesis process is a two-step procedure that involves specification detection and generation. 
In the first step, the text is examined to ascertain the presence or absence of a specification. 
If a specification is detected, in the second step the specification is synthesized. Figure~\ref{fig:model} shows the neural network model for \scheme.
Specification detection is a binary classification problem, where a BERT encoder is utilized. An encoder is a sequence of layers to convert the input texts in a hidden vector $h_c$. A special token $[CLS]$ is added at the beginning of the text to generate $h_c$
of the entire sequence for classification. This hidden state is subsequently passed to the decoder. A decoder is a sequence of layers to produce the final output from $h_c$.
The softmax layer in the decoder produces the binary prediction of the presence or absence of a specification in the input text. Basically, the specification classification process can be described as Equation~\ref{eq:bert_hidden}-\ref{eq:decoder_pred_class},
where $X=\{x_{CLS}, x_1, x_2,..., x_n\}$ is the tokenize set of input text, $x_{CLS}$ is the special token, $W$ is the weight matrix for the custom decoder and $c$ is the expected output prediction. We fine-tune $W$ to maximize the log probability of the correct class.
For specification classification tasks, BERT takes the final hidden state $h_\textsubscript{BERT}$ of the first token $[CLS]$ as the representation of the whole sequence. A simple softmax layer-based classifier is added as the custom decoder on the top of BERT encoder to predict the probability of label c.

\begin{equation}
    h_\textsubscript{BERT} = f\textsubscript{BERT}(X)
    \label{eq:bert_hidden}
\end{equation}

\begin{equation}
    \boldsymbol{h_c} = f_1(W_1 h_\textsubscript{BERT})
\end{equation}

\begin{equation}
    p(c|\boldsymbol{h_c}) = \text{softmax}(W_2\boldsymbol{h_c})
    \label{eq:decoder_pred_class}
    \vspace{0.5cm}
\end{equation}

For the specification synthesizing process, we use the same hidden state $h_c$  and pass it to a different sequence generation decoder. This decoder synthesizes the specification according to the pattern defined in Table~\ref{tab:spec_pat}. The synthesizing decoder is an LSTM-based~\cite{lstm} recurrent neural network defined as Equation~\ref{eq:decoder_synthesizer}. It takes $h_c$ as the hidden state and a start token to start the synthesis process and construct the specification as it goes.

\begin{equation}
    h_0 = \boldsymbol{h_c}
\end{equation}

\begin{equation}
    o_i = \text{LSTM}(h_{i-1}, o_{i-1})
    \label{eq:decoder_synthesizer}
\end{equation}

\begin{equation}
    o_\textsubscript{Specification} = \{o_1,...,o_m\}
    \vspace{0.5cm}
\end{equation}

A potential argument in favor of utilizing a single decoder for both the specification classification and generation tasks may be put forth. However, given that the majority of texts does not contain specifications and the generation of specifications is a complex task in comparison to binary classification, we found in our study that the use of two separate decoders yield better results in classifying and generating specifications within a text.

In our study, we categorize specifications into different categories. We can predict different classes of these categories with the same method. Upon detection of a specification, a decoder with a softmax layer having the desired number of classes can be used to classify different categories. A detailed analysis of the result is presented in the results section for this.

\subsubsection{Loss Function}

In the context of specification detection, a binary classifier is utilized to determine whether a given text contains any specifications or not. 
For this, we use weighted cross entropy loss function~\cite{arumugam2018hands}
defined by Equation~\ref{eq:loss_CE} where $M$ is the number of classes, $\log$ is natural log, $y$ is binary indicator (0 or 1) if class label $c$ is the correct classification for observation $o$, and $p$ is the predicted probability observation $o$ is of class $c$.

\begin{equation}
    \mathscr{L} = -\sum_{c=1}^{M} w_c y_{o,c}\log(p_o, c)
    \label{eq:loss_CE}
    \vspace{0.3cm}
\end{equation}
Weighted cross entropy loss is a commonly used loss function in classification problems when dealing with imbalanced datasets \cite{DBLP:journals/corr/abs-1708-02002, DBLP:journals/corr/HeZRS15}. 
This method assigns a higher weight to the minority class samples to improve the performance of the model. 
The weight is determined based on the distribution of the minority class samples in the dataset. 
By assigning a higher weight
to this class samples, the model is encouraged to focus more on correctly classifying these. 
Weighted cross entropy loss has been shown to be effective in improving the performance of models on imbalanced datasets, particularly in text classification tasks. 
Its usage can lead to a more accurate and reliable classification of the minority class.
In our specific dataset, it has been observed that the number of texts containing specifications is significantly lower than those that do not contain any specifications, leading to an imbalanced dataset. In order to address this issue, we have utilized weighted cross entropy loss function. The weight is determined based on the distribution of specification-containing texts in the training set.


%% file: sub_results.tex

In total, we collected 300 specifications from various sources, including 10 software manuals and other resources. The majority of the specifications were extracted from two sizable software manuals, due to their extensive page count. We categorize the software manuals into 3 groups according to their software types.
From Table~\ref{tab_data_man_desc}, first 2 softwares are categorized as Database software, next 4 as Distributed System and last 4 are categorized as Proxy. 

The quantity of available specifications are limited, as evidenced by prior research~\cite{pracExtractor} too . Given that the dataset used in the previous study is not publicly accessible, we collected same number of specification as them.
A deep learning based network requires a moderate number of training examples in order to train. 
However, given the lower number of available specification, it becomes challenging to create training and testing sets out of them.
Therefore we generate synthetic training examples by sampling 50 real specifications from our collected set and create the training set. 
Since we know the specification of these random samples, we put random text inside them to create synthetic samples. That way we create our training set that consists of 3000 samples in total. Thus, we were able to create a good amount of samples for the training and we use rest 250 samples as the testing set to evaluate.

In our model, we used a pretrained BERT model available online at 
https://huggingface.co/. 
Although, the BERT model is large in nature, since this is pretrained, we do not need to train them from scratch. 
We design our own decoder neural network for our own task. 
We use a feed forward neural network consisting of 2 hidden layer of 50 neurons each with a softmax function on top for specification detection. 
Weighted cross entropy loss is used for this decoder. 
To generate specifications, we use an LSTM-based recurrent neural network with 20 hidden units. 
Given the limited number of training examples, our designed neural network demonstrates sufficient capacity to achieve better prediction performance.
We train our model for 100 epochs until the training converges.

\begin{table*}[htbp]
\begin{tabular}{lcccccc}
\toprule
\multirow{2}{*}{Software Type}      & \multicolumn{3}{c}{PracExtractor} & \multicolumn{3}{c}{\scheme}   \\ \cline{2-7}
          & Precision   & Recall  & F1 Score  & Precision & Recall & F1 Score \\ \hline 
Database & 0.84          & 0.58      & 0.68        & 0.95        & 0.90     & 0.92       \\ \hline 
Distributed System & 0.79          & 0.50      & 0.61        & 0.90        & 0.75     & 0.82       \\ \hline
Proxy & 0.81          & 0.52      & 0.64        & 0.92        & 0.79     & 0.85       \\ \hline
{\bf Total} & 0.81          & 0.54      & 0.65        & 0.92        & 0.81     & 0.86       \\ \hline
\end{tabular}
\caption{\scheme's specification synthesis ability compare to PracExtractor}
\label{tab:result_baseline}
\end{table*}

\subsection{Demonstration of Synthesis Ability}

We compare our work with state-of-the-art specification synthesis tool named PracExtractor~\cite{pracExtractor}. PracExtractor uses Universal Dependency algorithm. This tool only work for {\it Simple} extraction type specification. On average a {\it Simple} extracted type specification is 3 tokens long. 
Therefore, given the short length of specifications, our model is capable of synthesizing them accurately once they are detected.
PracExtractor does not have any detection mechanism. Therefore, when they are able to synthesize a specification they are counted as successfully detected.

Table~\ref{tab:result_baseline} shows the specification detection performance comparison between PracExtractor and \scheme.
It showed result with three different software groups - Database, Distributed System and Proxy. PracExtractor performs poorly compared to \scheme in all three cases. In case of Database software, \scheme has higher Precision $0.95$ compared to $0.84$ Precision of PracExtractor. The Recall score is $0.90$ which is significantly better than Recall for PracExtractor $(0.58)$. Consequently F1 score for \scheme $(0.92)$ is significantly better than that of PracExtractor ($0.68$)
For Distributed System software type, \scheme has better Precision, Recall and f1 score compared to that of PracExtractor. In this case, Precision for \scheme is $0.90$ vs Precision of PracExtractor is$ 0.79$, Recall for \scheme is $0.75$ compared to $0.50$ Recall value for PracExtractor. In case of F1 Score, \scheme has higher value $0.82$ compared to the state of the art PracExtractor model $(0.61)$
In case of Proxy applications, we see similar trends in the result. Precision for \scheme is $0.92$ and Precision for PracExtractor is $0.81$ only. Recall is also much better for \scheme $(0.79)$ compared to Recall value for PracExtractor $(0.52)$. So F1 score for Proxy type software is higher in \scheme $(0.86)$ compared to F1 score of PracExtractor $(0.65)$. 

For combined results of all three types of software, we can see that \scheme has higher Precision $(0.92)$ compared to PracExtractor that has $(0.81)$ Precision only. This indicates that $92\%$ of the predicted relevant results were accurate for \scheme while only $81\%$ of the predicated relevant results were accurate for PracExtractor. \scheme has Recall value of $0.81$ compared to a mere $0.54$ Recall value for PracExtractor. Which signifies that \scheme could correctly identify $81\%$ of the relevant cases where PracExtractor could only correctly identify $54\%$ showing a huge improvement ($1.5\times$) in detecting relevant cases. Overall, \scheme has a F1 score of $0.86$ compared to F1 score of $0.65$ by PracExtractor which clearly explains that \scheme performed much better than the existing state of the art model.

\begin{table}[htbp]
\begin{tabular}{lccc}
\hline
              & True Positive & False Positive & False Negative \\ \hline
PracExtractor & 82\%          & 18\%           & 73\%           \\ \hline
\scheme       & 94\%          & 6\%            & 21\%            \\ \hline
\end{tabular}
\vspace{0.5cm}
\caption{Confusion matrix of specification detection capabilities}
\label{tab:result_false_positive}
\end{table}

Table~\ref{tab:result_false_positive} shows confusion matrix for specification detection between \scheme\ and PracExtractor. It shows \scheme provides better true positive and false positive compare to PracExtractor. In terms of true positive it is 12\% better in prediction. PracExtractor gives much high false positive (3x) prediction compare to \scheme. In terms of false negative detection, \scheme\ gives 21\% compare to 73\% of other tool.

\begin{table}[htbp]
\begin{tabular}{lcccc}
\hline
\multirow{2}{*}{Extraction Type} & \multicolumn{3}{c}{Detection}                                                             & \multirow{2}{*}{Generation} \\ \cline{2-4}
                                 & \multicolumn{1}{l}{Precision} & \multicolumn{1}{l}{Recall} & \multicolumn{1}{l}{F1 Score} &                             \\ \hline
Simple                           & 0.92                            & 0.81                         & 0.86                           & 100\%                          \\ \hline
Complex\textsubscript{Single}    & 0.81                            & 0.70                         & 0.75                           & 87\%                          \\ \hline
Complex\textsubscript{Multi}     & 0.73                            & 0.64                         & 0.68                           & 83\%                          \\ \hline
\end{tabular}
\vspace{0.5cm}
\caption{\scheme's Synthesis capability across different specification extraction type}
\label{tab:result_extraction_type}
\end{table}

Table~\ref{tab:result_extraction_type} demonstrates the specification synthesis results across different specification extraction type. The F1 score for specification detection in {\it Simple} extraction type is 0.86 where the maximum F1 score is 0.75 for the {\it Complex} category. The Precision in {\it Simple} extraction type is 0.92 that implies it is very good at detecting specification. Among {\it Complex} type {\it Complex\textsubscript{Single}} can detect specification with Precision of 0.81 where {\it Complex\textsubscript{Multi}} can detect specification with a Precision score of 0.73. However, the later category has higher F1 score than that of other. In terms of specification synthesis, 100\% specification can be synthesized in {\it Simple} category given they are correctly detected. Since, when the model is accurate at detecting specification the chances of synthesize a specification is higher. Moreover, in {\it Simple} cases average synthesized specification length is 3. Therefore, it is able to achieve 100\% accuracy in terms of specification generation. On the other hand, {\it Complex\textsubscript{Single}} synthesize 4\% more specification compare to {\it Complex\textsubscript{Multi}}.

\begin{table}[hbtp]
\begin{tabular}{lccc}
\toprule
\textbf{Category} & \textbf{Precision}  & \textbf{Recall} & \textbf{F1 score}\\
\midrule
 Quantitative & 0.95 & 0.82 & 0.88 \\ \hline
 Utilization & 0.8 & 0.89 & 0.84 \\ \hline 
 Interrelation & 0.8 & 0.89 & 0.84 \\ \hline
 Attribute & 0.95 & 0.90 & 0.92 \\ \hline
 Generic & 1.0 & 0.71 & 0.83 \\
 \bottomrule
\end{tabular}
\vspace{0.5cm}
\caption{\scheme Synthesis ability across different specification categories}
\label{tab:res_pred_categories}
\end{table}

\subsection{Performance of Specification Categorization}

Table~\ref{tab:res_pred_categories} shows Precision, Recall and F1 score across different categories of specifications - namely for Quantitative, Utilization, Interrelation, Attribute, Generic categories. For Quantitative category, \scheme had a Precision of $0.95$ and Recall of $0.82$ which gave a F1 score of $0.88$. For both Utilization and Interrelation categories, Precision value was $0.8$ while Recall was 0.89 and F1 score was 0.84. For Attribute category, \scheme performed better than previous categories with very high Precision of $0.95$ and Recall value of $0.9$ which gave the highest F1 score of $0.92$ across any categories. However, the highest Precision was achieved for Generic categories with $1.0$ value that is all the predicted relevant results were accurate for \scheme. However, for Generic category Recall value was a bit lower than other categories with 71\% of the relevant cases detected. So, F1 score was lower than other categories with a value of $0.83$

\subsection{Characterization of Models}

\begin{table}[hbtp]
\begin{tabular}{lccc}
\toprule
\textbf{Language Model} & \textbf{Precision}  & \textbf{Recall} & \textbf{F1 score}\\
\midrule
 BERT\textsubscript{Tiny} & 0.91 & 0.82 & 0.86 \\ \hline
 GPT\cite{gpt} & 0.91 & 0.80 & 0.85 \\ \hline 
 BERT & 0.92 & 0.81 & 0.86 \\
 \bottomrule
\end{tabular}
\vspace{0.5cm}
\caption{Model characterization with pre-trained language models}
\label{tab:res_pred_decoder}
\end{table}

We have used three different pretrained models and collected their Precision, Recall and F1 score. As we can see in Table~\ref{tab:res_pred_decoder}, BERT~\cite{bert} have the highest Precision among the models we have tested. BERT has Precision value of $0.92$, compared to Precision value of $0.91$ for both BERT\textsubscript{Tiny} and GPT~\cite{gpt} models. In case of identifying the fraction of relevant items that can be retrieved, i.e. the Recall score, BERT\textsubscript{Tiny} performs slightly higher $0.82$ than the BERT model $0.81$. GPT model has the lowest Recall score of $0.80$ among these three models. Combining these two score, the harmonic mean of Precision and Recall, known as F1 score, shows that both BERT and BERT~\textsubscript{Tiny} models perform better with F1 score of $0.86$ compared to the GPT model which has a F1 score of $0.85$. However, the performance of each of these three models is not significantly different from one another.



%% file: sub_rel.tex
\emph{Software configuration and specification extraction with NLP.} 
Numerous studies have addressed the challenge of effectively diagnosing and solving software configuration problems. 
One line of research focuses on using static analysis techniques to identify configuration errors before they result in system failures \cite{feamster2005detecting, rabkin2011static, nadi.icse.2014}. Other works, such as \cite{huang2018capturing} and \cite{attariyan2012x}, aim to enhance system observability to detect configuration errors in situ. Some approaches propose proactive methods to detect and troubleshoot customer issues, such as \cite{jin2010nevermind} and \cite{xu2016early}, which introduce early detection systems to prevent configuration errors from causing significant damage. Online error detection systems based on context have been proposed in \cite{yuan2011context}, while \cite{zhang2014encore} proposes a misconfiguration detection system based on system environment and correlation information.
Shared knowledge or event traces have been used to diagnose misconfigurations in \cite{agarwal2009netprints} and \cite{yuan2006automated}, respectively. An automated troubleshooting approach based on dynamic information flow analysis is presented in \cite{attariyan2010automating}, while \cite{quinn2016jetstream} proposes a parallelized approach to information flow queries. Precomputing approaches to possible configuration error diagnoses have been proposed in \cite{rabkin2011precomputing}. \cite{wang2004automatic} proposes a troubleshooting approach based on peer pressure, while \cite{wang2004strider} suggests a state-based approach to change and configuration management. On the other hand, \cite{whitaker2004configuration} proposes a search-based approach to configuration debugging, and \cite{zhang2013automated} introduces an automated diagnosis system for configuration errors.
In addition to technical approaches, some papers emphasize the importance of not blaming users for configuration errors and instead focusing on developing better tools and processes to prevent them, such as in \cite{xu.sosp.2013}. Probabilistic approaches have also been proposed to learn configuration file languages and identify and diagnose configuration problems, as demonstrated in \cite{santolucito2016probabilistic}. Similarly, \cite{configV} presents an association rule learning-based approach to synthesize configuration file specifications from a set of example configurations. Also, \cite{su2007autobash} proposes a causality analysis-based technique to identify the root cause of configuration errors and automate the configuration management process.

One particularly influential work in this area is the PracExtractor \cite{pracExtractor}, which employs natural language processing to analyze specific configurations and convert them into specifications to identify potential system admin flaws. However, PracExtractor has limitations, such as inflexibility and low generalization ability due to the specific format required for accurate extraction. Additionally, PracExtractor struggles with large paragraph settings, which our SpecSyn system seeks to improve upon. Other research efforts have also explored methods for inferring specifications from text \cite{tan2007icomment, tan2011acomment, pandita2012inferring, zhai2016automatic, zhong2009inferring, zhou2017analyzing, wong2015dase, enck2009configuration, loo2005declarative, chen2010declarative}, but they too have limitations. Our work builds on these prior efforts by leveraging their insights to provide a more accurate and effective model for specification extraction.

\emph{Specification extraction using Knowledge Base.} The use of a Knowledge Base (KB) has been explored in prior research for specification extraction, as seen in ConfSeer\cite{potharaju2015confseer}. However, similar to PracExtractor, this approach has its limitations. ConfSeer relies on a KB to analyze potential configuration issues, which can make it feel more like a search engine. While this approach has its benefits, it can also limit flexibility. SpecSyn aims to address this limitation by analyzing unstructured data to provide a greater variety of specifications. Other systems have been proposed to assist with finding configurations, such as\cite{wang2004automatic, whitaker2004configuration}, but they come with their own issues, such as high overheads and requirements for large datasets. Associated rule learning has been used in previous studies to address dataset issues\cite{mabroukeh2010taxonomy, ayres2002sequential, han2007frequent,langley1995applications}.


\emph{BERT based extraction and NLP} 
In contrast to previous studies, we employed and fine-tuned the BiDirectional Encoder Representations and Transformers model, or BERT\cite{bert}, to gain a better understanding and improve the training of our dataset. BERT is a new natural language processing model that offers a more in-depth and refined fine-tuning technique. However, there are some limitations associated with using BERT. As it was developed in 2018, it is still a relatively new model and may not be fully developed with regard to training sets. Additionally, it can be expensive to train and result in slower training times due to its many weights\cite{projectpro_2022}. Despite these limitations, BERT does an excellent job of processing specific input and producing output with new specifications, providing great flexibility as it eliminates the need for a KB as used in ConfSeer and can expand upon the findings of the PracExtractor system.


In recent times, deep learning is heavily used to address diverse system-related issues, such as program synthesis~\cite{netsyn, genesys}, partial program correction~\cite{gupta2017deepfix}, bug fixing~\cite{li2022dear} etc.
Several studies have employed natural language processing (NLP) to inspect and analyze software configurations, particularly in the context of security analysis\cite{stockle2022automated}. This work focused on analyzing two security systems, CIS and Siemens, and trained two models, LDA and BERT, which is the model used in our study. However, this study had certain limitations, such as the narrow focus on only two types of security systems, which could limit the diversity of the training data. Moreover, the study found that the BERT model, according to their metric of measurement, was not accurate enough to detect misconfigurations. Nevertheless, this study provided a useful baseline for identifying configuration issues using NLP and made their dataset publicly available on Kaggle, facilitating further research in the field.


Recent studies have also investigated software configuration and misconfiguration using state analysis techniques\cite{li2021software}. In\cite{li2021software}, the authors developed ConfDetect, a system that analyzes log files and ranks them based on specific criteria, which can effectively predict misconfiguration errors. The system also utilizes NLP to extract logs. The authors reported substantial accuracy in diagnosing configuration errors. However, the system was only tested on three different systems, whereas our study has more variety in terms of data testing. Additionally, ConfDetect utilizes a knowledge base to detect configuration errors, which could limit its flexibility. Despite these limitations, the study provides valuable insights into finding proper software misconfiguration.




%% file: sub_future.tex
In this section, we aim to highlight some potential observations and discuss them in detail.

\emph{Why use LSTM base decoder for generating specifications while Transformer base decoder is considered as the state-of-the-art?}
The Transformer architecture offers advantages over the LSTM-based architecture in two scenarios: first, when dealing with a tremendously large dataset (i.e. 100GB) that requires parallel training without any previous recurrence dependency, and second, when handling very long sequences~\cite{transformer}. In our specific case, the dataset that we use is comparatively very small, and the sequence to be generated is also relatively short. As a result, the utilization of a Transformer-based decoder over an LSTM-based decoder will not bring any benefits.


\emph{Why applying data composition instead of passing original text?}
Data composition helps the model to generalize better. 
While it is feasible to train a model without data composition and achieve similar performance results, doing so would require a larger quantity of data and can be left for future exploration.

\emph{How long sequence has been used to synthesize specification?}
In this study, we limit ourselves to check one sentence for {\it Simple} extraction type and two sentences for {\it Complex} extraction type. 
Our findings indicate that two sentences adequately cover the majority of {\it Complex} extraction type specifications. Furthermore, our sample sentences have an average length of approximately 150 characters, which implies that we check a maximum of around 300 characters for {\it Complex} extraction. 
We also train our model with a maximum of two sentences long samples. As a result, the model's performance will be poor for longer sequences than it is trained on. Hence, it would require more data samples and a potentially larger parametric model architecture to overcome this issue. Therefore, we keep this as a future work. It would be valuable to investigate very long sequences and conduct a sequence length sensitivity analysis.

Furthermore, regarding comments written in software source code, it is possible to parse a greater number of software programs. Additionally, examining the neighboring source code can provide a better understanding of the context of the comments. For extracting specification from other online discussion forums, a more exhaustive search could be conducted to mine a larger set of specifications. However, this is beyond the scope and context of our current project.


%% file: sub_conc.tex
In this paper we introduces a deep learning-based framework for automatic specification synthesis using a large language model. To the best of our knowledge, this is the {\bf first} work to utilize a large language model to understand natural language context and synthesize specifications.
We formulate specification synthesis task as a sequence learning problem and integrate BERT, a pre-trained large language model for this purpose. 
Our proposed framework \scheme\ outperforms prior state-of-the-art by 21\% in terms of F1 score. This work opens up new direction to synthesize specification and can be extended for other similar system-related works as well.